\documentclass{PoS}

\usepackage{amsmath}
\usepackage{caption,subcaption}

\usepackage{comment}

\title{$(1+1)$-d $U(1)$ Quantum link models from effective Hamiltonians of dipolar molecules}

\ShortTitle{QLMs from dipolar molecules}

\author{\speaker{Jiayu Shen}\thanks{E-mail: {jiayus3@illinois.edu}} \thanks{These authors contributed equally to this work.}, Di Luo\footnotemark[\value{footnote}], Michael Highman, Bryan K.\ Clark, Brian DeMarco, Aida X.\ El-Khadra, Bryce Gadway \\
        Department of Physics, University of Illinois, Urbana, IL 61801, USA\\
        }

\abstract{
We study the promising idea of using dipolar molecular systems as analog quantum simulators for quantum link models, which are discrete versions of lattice gauge theories. In a quantum link model the link variables have a finite number of degrees of freedom and discrete values. We construct the effective Hamiltonian of a system of dipolar molecules with electric dipole-dipole interactions, where we use the tunable parameters of the system to match it to the target Hamiltonian describing a $U(1)$ quantum link model in $1+1$ dimensions.
}

\FullConference{37th International Symposium on Lattice Field Theory - Lattice2019\\
		16-22 June 2019\\
		Wuhan, China}

\begin{document}

\section{Introduction}
The idea of quantum simulations is to study quantum field theories using quantum hardware \cite{Feynman:1981tf}.
Lattice gauge theories (LGT) are
a particularly interesting target for quantum simulations. Both, digital \cite{Martinez_2016} and analog
\cite{Zohar-report-15}
quantum simulations show promise for studies of the real-time evolution of LGTs in a Hamiltonian formalism.  
Since, generally, the number of available quantum states in an experimental system is finite, a truncation scheme (or other approximation) is needed to reduce the infinite degrees of freedom in a gauge theory to a finite number.  
Quantum link models (QLMs) \cite{Chandr-QLM} provide a viable reformulation of LGTs for quantum simulations. Experiments for analog quantum simulations of QLMs using Hubbard models \cite{Banerjee_2012} and Rydberg atoms \cite{surace2019lattice} have been proposed. In this work, we propose quantum simulation experiments of QLMs using dipolar molecules at fixed positions.

\section{$(1+1)$-d $U(1)$ quantum link model}

We consider the massive lattice Schwinger model \cite{PhysRev.128.2425, COLEMAN1976239}, i.e., the $(1+1)$-d $U(1)$ LGT. With staggered fermions $\psi_x$ and in temporal gauge $A_0 = 0$, the Hamiltonian takes the form 
\begin{equation}
H = - w \sum _ { x } \left[ \psi _ { x } ^ { \dagger } U _ { x , x + 1 } \psi _ { x + 1 } + \psi _ { x + 1 } ^ { \dagger } U _ { x , x + 1 } ^ { \dagger } \psi _ { x } \right] + m \sum _ { x } ( - 1 ) ^ { x } \psi _ { x } ^ { \dagger } \psi _ { x } + \frac { g ^ { 2 } } { 2 } \sum _ { x } E _ { x , x + 1 } ^ { 2 }
,
\label{eq:qlm}
\end{equation}
where $w > 0$ is the hopping parameter, $m$ is the fermion mass, and $g$ is the gauge coupling. The gauge field $A_{x, x+1} := (A_1)_{x, x+1}$ is formulated in the compact form $U_{x, x+1} = \exp (i a g A_{x, x+1})$, so that the electric field, which is proportional to the conjugate momentum of $A_{x, x+1}$, has a discrete but infinite spectrum, $E_{x, x+1} = - \frac{i}{a g} \frac{\partial}{\partial A_{x, x+1}} \in \mathbb{Z}$ \cite{Kogut-LGTreview-1979}.  In $(1+1)$ dimensions staggered fermions have only one Dirac component.
Each fermion site has two possible quantum states: occupied and empty. However, the degrees of freedom for the gauge links are infinite. The commutation relations on a link, $[ E_ { x , x + 1 } , U_ { x , x + 1 } ] = U_ { x , x + 1 } $,  $[ E_ { x , x + 1 } , U_ { x , x + 1 } ^ { \dagger } ] = - U_ { x , x + 1 } ^ { \dagger } $,  $[ U_ { x , x + 1 } , U_ { x , x + 1 } ^ { \dagger } ] = 0$, are reminiscent of the $SU(2)$ algebra $[ S^3, S^{+} ] = S^{+} $, $[ S^3 , S^{-} ] = - S^{-}$, $[S^{+}, S^{-} ] = 2 S^3$ except for the last commutator. Imposing $[ U_ { x , x + 1 } , U_ { x , x + 1 } ^ { \dagger } ] = 2 E_ { x , x + 1 }$, instead of $[ U_ { x , x + 1 } , U_ { x , x + 1 } ^ { \dagger } ] = 0$, changes this theory to a quantum link model. A consequence of this modification is that the spectrum of the electric flux becomes finite $E_{x, x+1} = \in \lbrace -S, -S + 1, \cdots, S - 1, S \rbrace$ for fixed $S = 1/2, 1, 3/2, \cdots$, where different values of $S$ correspond to different QLMs. In summary, QLMs are ideal for quantum simulations, because their Hilbert spaces are finite. 
Here, we focus on $S = 1/2$ and $S = 1$ $U(1)$ QLMs. In particular, QLMs with $S \ge 1$ exhibit Schwinger effect \cite{PhysRev.82.664} or string breaking \cite{Banerjee_2012, Hauke_2013} known in the Schwinger model. As another motivation, a $(d+1)$-dimensional quantum link model can be regarded as a $D$-theory of the $d$-dimensional Euclidean LGT through a dimensional reduction in a fictitious Euclidean dimension \cite{Schlittgen_2001}.

Analogous to its LGT counterpart, the Gauss' law operator of a QLM evaluated at any $x$, $\widetilde { G } _ { x } = \psi _ { x } ^ { \dagger } \psi _ { x } - E _ { x , x + 1 } + E _ { x - 1 , x } + \frac { 1 } { 2 } [ ( - 1 ) ^ { x } - 1 ]$ must satisfy $\widetilde { G } _ { x } |\mathrm{phys}\rangle = 0$ in the absence of background fields. Here $|\mathrm{phys}\rangle$ denotes any physical state, $|\mathrm{phys}\rangle \in \mathcal{H}_{\mathrm{phys}}$. This Gauss' law condition is imposed by starting with a physical initial state $|\Psi (t = 0) \rangle \in \mathcal{H}_{\mathrm{phys}}$, then the fact that $\widetilde { G } _ { x }$ is conserved, i.e.\ $[H, \widetilde { G } _ { x }] = 0$, ensures that the time-evolved state $|\Psi (t) \rangle = e ^{-i H t}|\Psi (t = 0) \rangle$ is guaranteed to stay in $\mathcal{H}_{\mathrm{phys}}$.

In the $S=1/2$ QLM, electric fluxes may form as strings, pointing in the left or right direction. These two orientations of electric flux are $C$ and $P$ conjugates of one another. Oscillations between these two directions of the string, referred to as string inversion \cite{Hauke_2013, surace2019lattice}, occur in real-time dynamics.

\section{Physics of dipolar molecules}

Dipolar molecules are molecules composed of two atoms and are known as a physical platform of quantum magnetism \cite{Wall_2015}. Under the envisioned experimental conditions, they can be thought of as rods with fixed lengths. The rotation of a dipolar molecule is quantum mechanically characterized by an orbital angular momentum $\mathbf{N}$ of which the eigenstates are $|N, m_N\rangle$ as usual. The rotor Hamiltonian for the dipolar molecule is $H_{\mathrm{rot}} = h B_{\mathrm{rot}} \mathbf{N}^2$ where $B_{\mathrm{rot}}$ is a rotational constant that depends on the molecule species. Here, we use only the $N = 0$ and $N = 1$ states. The $N \ge 2$ states have larger energes and are off-resonance for our purpose. For simplicity, we denote the $N=0,1$ states as $|a\rangle := |0, 0\rangle$, $|b\rangle := |1, -1\rangle$, $|c\rangle := |1, 0\rangle$, $|d\rangle := |1, 1\rangle$.

In the absence of an external field, a dipolar molecule is isotropic. However, a strong uniform magnetic field can impose a specific direction for the basis of angular momentum eigenstates, namely, the $z$-axis of spherical harmonics or the quantization axis \cite{Neyenhuis-Polarizability-12}. When the geometrical configuration of molecules already sets a direction, then the system itself becomes anisotropic and the quantization axis has relative angles with respect to the geometric configuration.

Two dipole molecules $i$ and $j$ at fixed positions have an electric dipole-dipole interaction ${ V } _ { i j }
	= ( 4 \pi \epsilon _ { 0 } )^{-1} [ { \mathbf { d } } _ { i } \cdot { \mathbf { d } } _ { j } - 3 ( { \mathbf { d } } _ { i } \cdot \hat { r } _ { i j } ) ( { \mathbf { d } } _ { j } \cdot \hat { r } _ { i j } ) ] / r _ { i j } ^ { 3 } $
where $\mathbf{d}_i$ and $\mathbf{d}_j$ are electric dipoles and $\mathbf{r}_{ij}$ is the separation from $i$ to $j$. $V_{ij}$ should be treated as a quantum operator whose matrix elements depend on the quantum states of the two molecules. Operators $\mathbf{d}_i$ and $\mathbf{d}_j$ act on two different molecules. Each molecule has a dipole selection rule that $\left\langle N ^ { \prime } , m _ { N } ^ { \prime } \left| \mathbf { d } \right| N , m _ { N } \right\rangle \ne 0$ only when $\Delta N = N' - N = \pm 1$ and $\Delta m_N = m_N' - m_N = 0, \pm 1$.

In the experiment, laser beams are applied to the dipolar molecules to alter the relative energy shifts between the $N = 0$ and $N = 1$ states \cite{Neyenhuis-Polarizability-12}. Each molecule is controlled with its own laser beam. As a result, the potential energies introduced by the laser beams are both position and state-dependent. In the Hamiltonian, the effects of the laser lights are $\sum_{i, N, m_N} \epsilon_{i, N, m_N} n_{i, N, m_N}$ where $N, m_N$ can take $a$, $b$, $c$ and $d$.

The molecular positions, external magnetic field, and laser beams are all experimentally adjustable, yielding tunable parameters which can be used to map the molecular system to the target QLM Hamiltonian.

\section{Mapping the dipolar molecular system to the target QLM}
\label{sec:mapping}

In order to interpret the experimental system as the target theory, a mapping from the dipolar molecular system to the QLM must be established. In our proposal, each dipolar molecule is identified with either a site or a link in the QLM, where the accessible molecular states are mapped to one-body states on the sites or links of the QLM. The experimental system is simply a chain of molecules, where, in our case, we choose molecules that are characterized as hard-core bosons \cite{lacroix2011introduction}. Hardcore bosons have strong repulsive hard-core potentials, so that each state can, at most, be occupied by one boson. This property enables their use as quantum spins on links, and, in $1+1$ dimensions, also as fermions on sites after a Jordan-Wigner transformation.

Each fermion site has two possible states ($|a\rangle$ and $|c\rangle$), where $|a\rangle$ is mapped to the occupied state and $|c\rangle$ is mapped to the empty state. The $|b\rangle$ and $|d\rangle$ states are made off-resonant by adjusting their energies such that they are dynamically inaccessible (or highly suppressed) to the molecules at the fermion sites.

Each link has $2 S + 1$ states. For an $S = 1/2$ QLM, $|b\rangle$ is mapped to the $S^3 = -1/2$ (spin down) state and $|d\rangle$ is mapped to the $S^3 = 1/2$ (spin up) state. $|c\rangle$ is made off-resonant. $|a\rangle$ is not directly used in the mapping but it plays a role dynamically and can appear in intermediate states of virtual processes that will be discussed in a later section. For an $S = 1$ QLM, $|d\rangle$ is mapped to the $S^3 = -1$ state, $|b\rangle$ is mapped to the $S^3 = 0$ state and $|c\rangle$ is mapped to the $S^3 = 1$ state. $|a\rangle$ again is not used directly in the mapping but plays a role in virtual processes.

\vspace{-0.7cm}
\section{Engineering the effective Hamiltonian}

The target QLMs contain one-body terms for the fermion mass ($m \sum_x (-1)^x \psi_x^\dagger \psi_x$) and the electric flux energy ($\frac{g^2}{2} \sum_x E^2_{x,x+1}$). In addition, there is a three-body interaction, the fermion hopping term, $-w\sum_x ( \psi^\dagger_x U_{x,x+1} \psi_{x+1} + h.c.)$, which involves two neighboring sites and the link in between them.
In the dipolar molecular system the rotor Hamiltonian and the laser lights provide one-body terms. However, the molecular Hamiltonian contains only two-body interactions in the form of dipole-dipole terms. In order to describe the three-body interactions in the QLM Hamiltonian, we must consider higher-order interactions in the molecular system. Here we use second-order perturbation theory to construct the quasi-degenerate effective Hamiltonian \cite{winkler2003spin}
, which acts on a quasi-degenerate subspace of the Hilbert space. 

We denote the quasi-degenerate subspaces by $\alpha$, $\beta$, $\gamma$ ... where we are interested in the effective Hamiltonian for one such subspace, say, $\alpha$. We use $m$, $l$, $n$ ... to denote the states in that subspace, which, in our case, are tensor product states of the individual molecular states, i.e., a sequence of $|a\rangle$, $|b\rangle$, $|c\rangle$ and $|d\rangle$ whose length equals the number of molecules. Since the states $m$ are quasi-degenerate, their energies, while $m$ dependent, vary by $\delta E$, which is small. However, the energy differences between two subspaces are large, $E_{m \alpha} - E_{n \beta} \sim \Delta \gg \delta E$ (independent of the values of $m$ and $n$). 

In our construction of the QLM Hamiltonian, $\alpha$ should target $\mathcal{H}_{\mathrm{phys}}$. Each state in $\alpha$ is a configuration of the molecular chain that obeys the Gauss' law after mapping to the QLM states. In the Hamiltonian of the dipolar molecular system, $H = H_0 + V$, we treat $V = \sum_{i < j} V_{ij}$ the dipole-dipole interactions, where $i$ and $j$ are molecular indices, as a perturbation. The Hamiltonian $H_0 = \sum_i ( H_{\mathrm{rot}, i} + H_{\mathrm{laser}, i} )$ is exactly solvable. The eigenstates of $H_0$ are tensor products of each molecule's angular momentum eigenstate and serve as the basis of performing the perturbation theory. Experimentally, $H_0$ can be adjusted to the desired form by tuning the laser beams.
At second order in perturbation theory, the matrix elements of the quasi-degenerate effective Hamiltonian for the space $\mathcal{H}_{\mathrm{phys}} = \alpha$ take the form
\begin{equation}
\begin{aligned}
	\left\langle m \left| H _ { \mathrm { eff } } ^ { \alpha } \right| n \right\rangle = & E _ { m \alpha } \delta _ { m, n } + \langle m, \alpha | V | n , \alpha \rangle \\ & + \frac { 1 } { 2 } \sum _ { l , \gamma \neq \alpha } \langle m , \alpha | V | l , \gamma \rangle \langle l , \gamma | V | n , \alpha \rangle \left[ \frac { 1 } { E _ { m \alpha } - E _ { l \gamma } } + \frac { 1 } { E _ { n \alpha } - E _ { l \gamma } } \right] + \cdots
	.
\end{aligned}
\label{eq:eff_ham}
\end{equation}
The zeroth-order terms $E _ { m \alpha } \delta _ { m, n }$ correspond  to $H_0$ projected to the subspace $\alpha$. The first-order terms $\langle m , \alpha | V | n , \alpha \rangle$ are matrix elements of the dipole-dipole interaction. Any two distinct states in $\alpha$ that satisfy Gauss' law must differ in at least three molecules. However, because $V$ is a two-body interaction, only matrix elements of $V$ between states that differ in two molecules are nonzero. We can therefore neglect the first-order terms in Eq.~(\ref{eq:eff_ham}).

The second-order term in Eq.~(\ref{eq:eff_ham}) describes the effects on states in $\alpha$ from states outside $\alpha$. The state $|l, \gamma \rangle$ is an intermediate state that appears only in the  virtual process $|n, \alpha \rangle \rightarrow |l, \gamma \rangle \rightarrow |m, \alpha\rangle$. The strength of this second-order perturbation is determined by the matrix elements of $V$ and the energy differences in $H_0$. We can tune the matrix elements of $V$ by varying the inter-molecular separations and adjust the energy differences in $H_0$ by tuning the lasers. By virtue of these tunable, experimental parameters, we can adjust the effective Hamiltonian so that it becomes equivalent to
$
- w \sum _ { x } ( \psi _ { x } ^ { \dagger } U _ { x , x + 1 } \psi _ { x + 1 } + h. c. )
$
with a constant hopping parameter $w$. 
Self-interactions also arise at second-order in perturbation theory, because the virtual process described above can involve the same initial and final state, i.e., $|n, \alpha \rangle \rightarrow |l, \gamma \rangle \rightarrow |n, \alpha\rangle$. However, self-interactions are diagonal and do not break Gauss' law. In numerical tests of the dynamics, we find that they do not significantly affect the important features. We can also use residual freedom in tuning the inter-molecular separations to further suppress the self-interactions. More details regarding the experimental set-up and the tuning of the parameters are given in \cite{luo2019framework}.

\section{Numerical tests}

\begin{figure}[!htb]
  \centering
  \begin{subfigure}[b]{0.46\linewidth}
    \centering\includegraphics[width=1.0\textwidth]{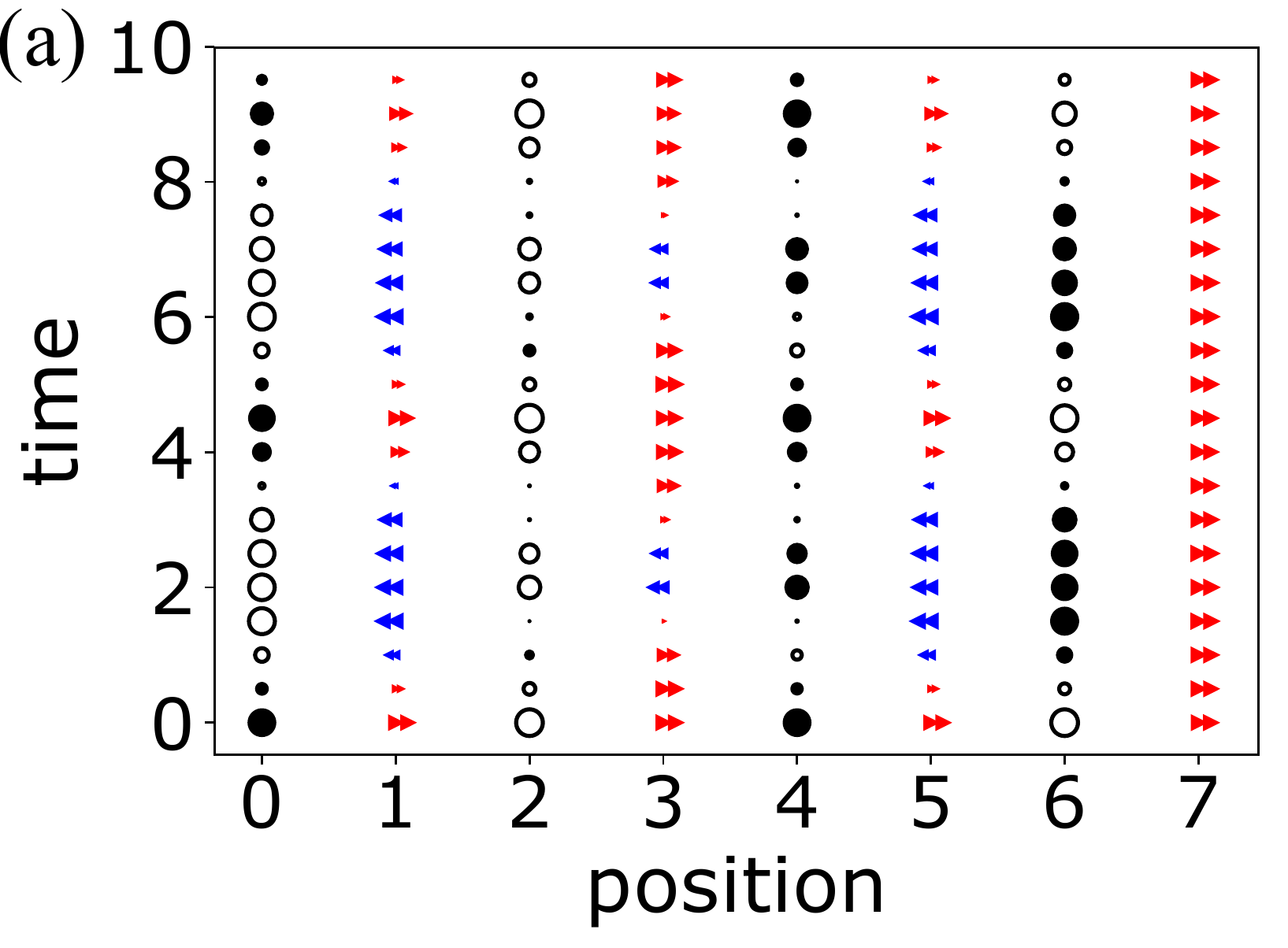}
    \phantomsubcaption\label{fig:string_inversion_1}
  \end{subfigure}
  \begin{subfigure}[b]{0.52\linewidth}
    \centering\includegraphics[width=1.0\textwidth]{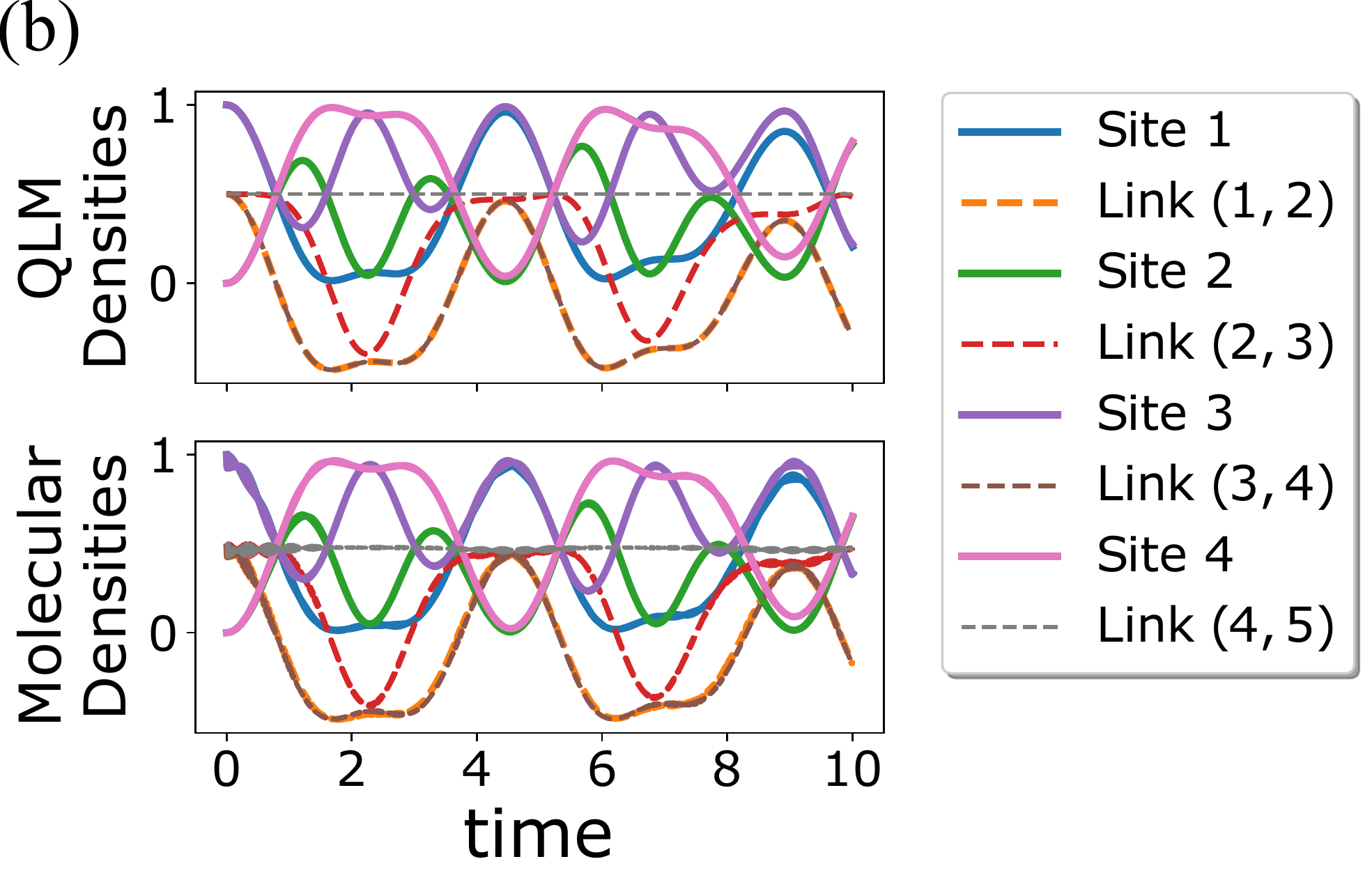}
    \phantomsubcaption\label{fig:string_inversion_2}
  \end{subfigure}
  
  \begin{subfigure}[b]{0.46\linewidth}
    \centering\includegraphics[width=1.0\textwidth]{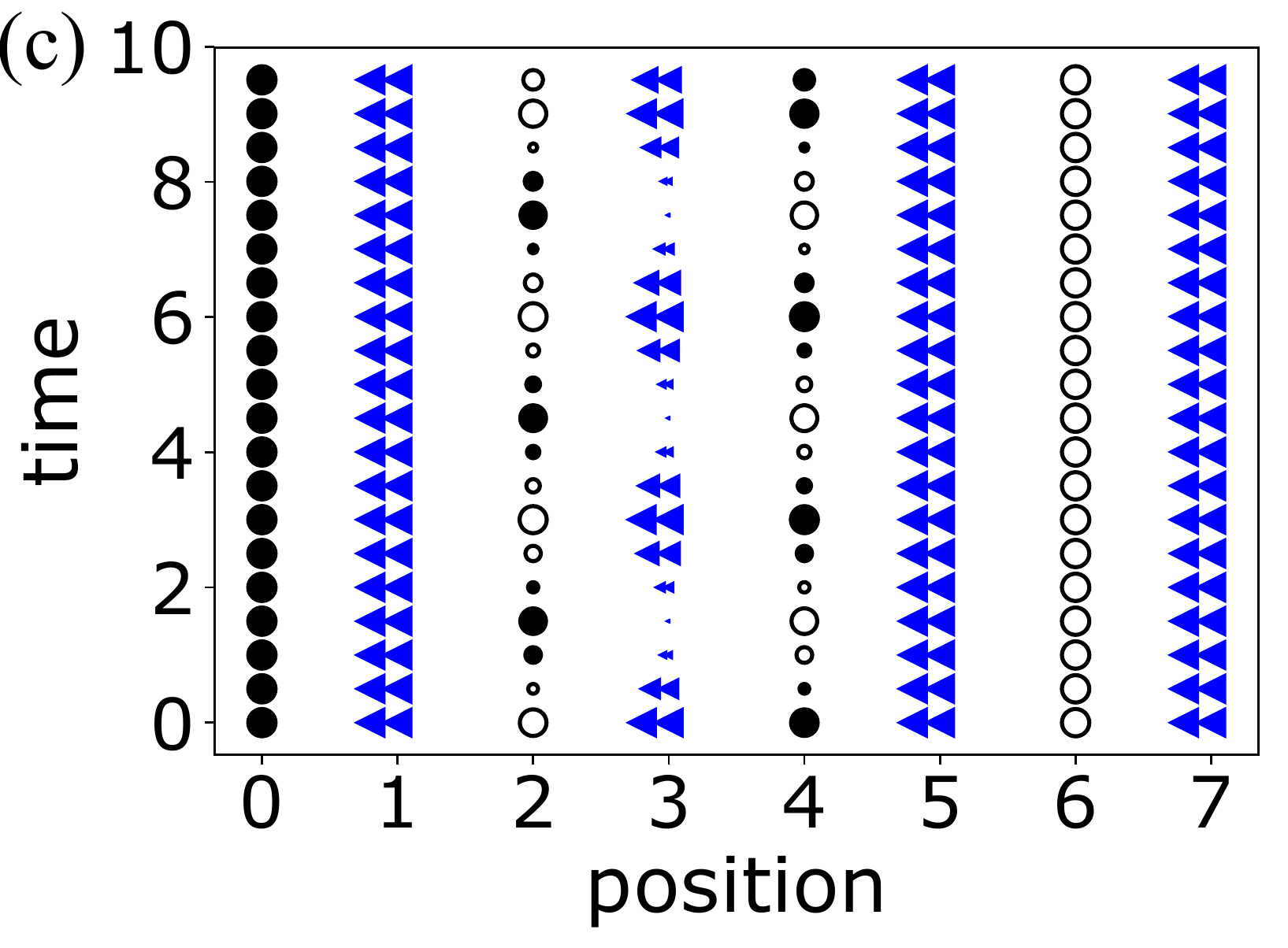}
    \phantomsubcaption\label{fig:string_breaking_1}
  \end{subfigure}
  \begin{subfigure}[b]{0.52\linewidth}
    \centering\includegraphics[width=1.0\textwidth]{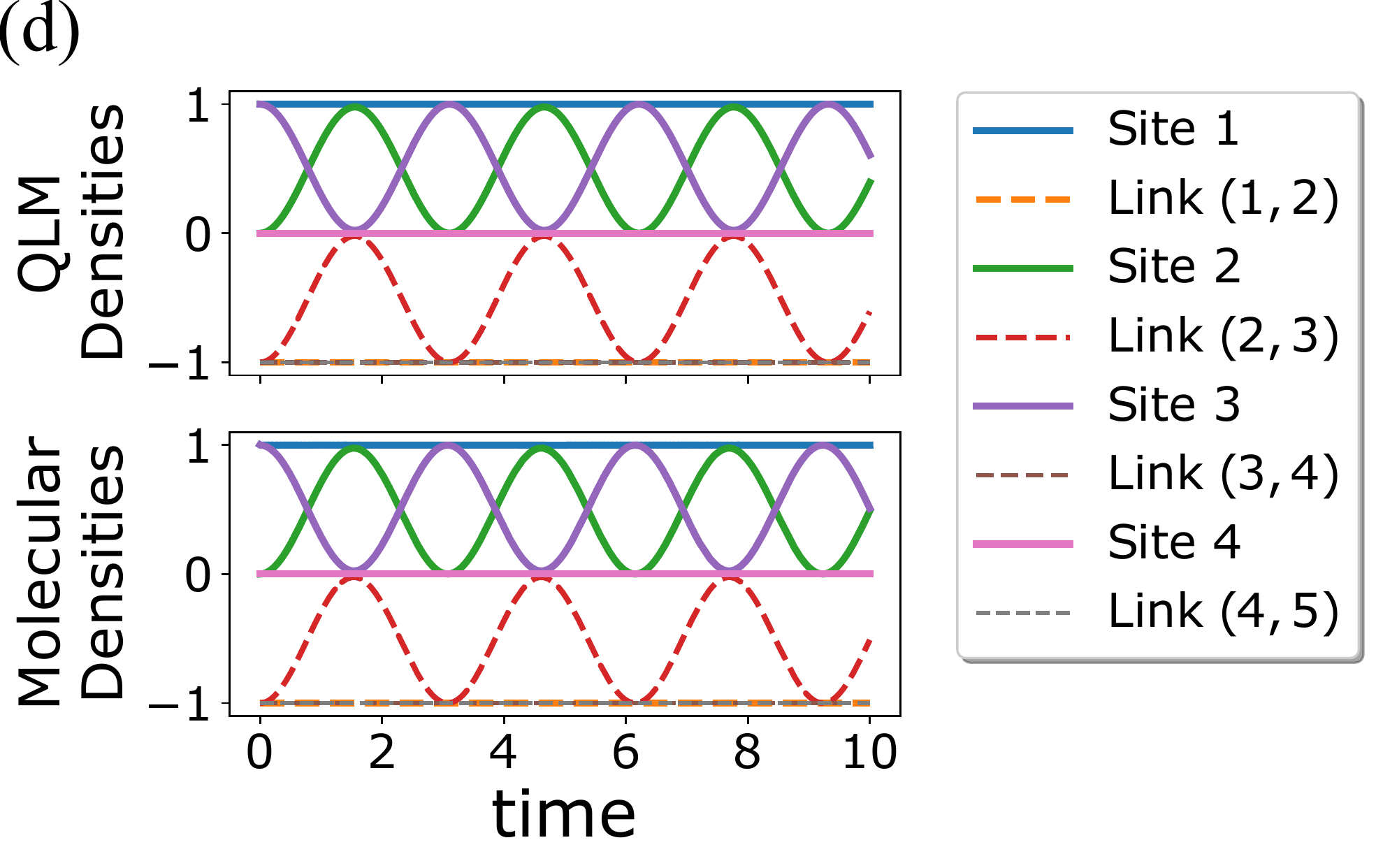}
    \phantomsubcaption\label{fig:string_breaking_2}
  \end{subfigure}\
  \caption{(\subref{fig:string_inversion_1}) and (\subref{fig:string_inversion_2}): $S = 1/2$ QLM at $m = 0.1 w$ in two unit cells with the open boundary condition. Time is in units of $1/w$. (\subref{fig:string_breaking_1}) and (\subref{fig:string_breaking_2}): $S = 1$ QLM at $m = 0.1 \times \sqrt{2} w$, $g^2 / 2 = 0.5 \times \sqrt{2} w$ in two unit cells with the open boundary condition. Time is in units of $1/(\sqrt{2} w)$.
  Time evolution of expectation values of fermion densities and electric fluxes in the dipolar molecular system after mapping to QLM states for (\subref{fig:string_inversion_1}) $S = 1/2$ (\subref{fig:string_breaking_1}) $S = 1$. Filled circles denote fermion occupied states and hollow circles denote fermion empty states. Red right-pointing arrows denote right-pointing electric fluxes and blue left-pointing arrows denote left-pointing electric fluxes. Full-sized symbols correspond to extrema of expectation values and non-full-sized symbols denote how close the expecation values are to the extrema.
  In (\subref{fig:string_inversion_1}), the direction of the electric flux string exhibits an oscillation over time, i.e., string inversion.
  In (\subref{fig:string_breaking_1}), the electric flux starts from a left-pointing string, but later on the string breaks modulo finite size effects \cite{Pichler-PRX}.
  With the parameters we use, the broken state and the recovered state oscillates from one another.
  Comparisons between the QLM link/site densities and dipolar molecular link/site densities as functions of time for (\subref{fig:string_inversion_2}) $S = 1/2$ (\subref{fig:string_breaking_2}) $S = 1$.
  The fermion density is rescaled to the interval $[0, 1]$. The electric flux ranges from (\subref{fig:string_inversion_2}) $-0.5$ to $0.5$ (\subref{fig:string_breaking_2}) $-1$ to $1$.
  }
  \label{fig:numerics_1}
\end{figure}

We use the method of exact diagonalization (ED) to test how well the molecular system describes the target theory, where we consider systems of up to two cells in this report. 
We start the two systems in the same initial state and time-evolve them with the two Hamiltonians, $| \Psi_{\mathrm{QLM}} (t) \rangle = e^{- i H_{\mathrm{QLM}} t} | \Psi_{0} \rangle$ and $| \Psi_{\mathrm{molecule}} (t) \rangle = e^{- i H_{\mathrm{molecule}} t} | \Psi_{0} \rangle$. We do not expect the two time evolutions to be identical even if we had perfect control over the experimental setup and parameters, because the effective Hamiltonian is based on second-order perturbation theory and includes terms only through order $V^2/\Delta$. 
Our results for the time evolution of site and link density expectation values and comparisons between dynamics of the QLM and molecular systems for $S=1/2$ and $S = 1$, respectively, are plotted in Figure \ref{fig:numerics_1}. More details of numerical tests can be found in \cite{luo2019framework}.

\section{Summary and outlook}

We propose a new approach for analog quantum simulations of quantum link models that uses dipolar molecules. We show that it is possible to simulate $S = 1/2$ and $S = 1$ QLMs in $1+1$ dimensions with the envisioned experimental set-up. Our studies of small systems with two unit cells reveal  that the effective Hamiltonian of the dipolar molecular system is able to numerically reproduce the dynamical phenomena of string inversion and string breaking in QLMs. 
These results open the possibility of realistic simulations of larger systems and experimental work to implement these analog simulations in the lab. In the future we plan to explore dipolar molecular systems as experimental platforms for quantum simulations of other lattice field theories.

\section*{Acknowledgments}

We thank P.\ Draper, Y.\ Meurice, and J.\ R.\ Stryker for discussions.
This work was supported in part by the U.S.\ Department of Energy under award No.\ DE-SC0019213 and by the National Science Foundation Graduate Research Fellowship Program under Grant No.~DGE–1746047 (M.H.).
J.S.\ is grateful for support from the UIUC Department of Physics and the Lattice 2019 organizers which enabled him to attend this conference. 

\bibliographystyle{JHEP}
\bibliography{main}

\end{document}